\documentclass[twocolumn,superscriptaddress,pre]{revtex4}

\usepackage{amsmath}
\usepackage{amsbsy}
\usepackage{mathtools}
\usepackage{bbm}
\usepackage{dcolumn}
\usepackage{enumitem}
\usepackage{graphicx}
\usepackage{float}

\newcommand{\defn}{\textit}
\newcommand{\var}{\mathop{\textrm{Var}}}
\newcommand{\cov}{\mathop{\textrm{Cov}}}

\begin{document}
\title{The friendship paradox in real and model networks}
\author{George T. Cantwell}
\email[Email address: ]{gcant@umich.edu}
\affiliation{Santa Fe Institute, 1399 Hyde Park Road, Santa Fe, New Mexico 87501, USA}
\affiliation{Department of Physics, University of Michigan, Ann Arbor, Michigan 48109, USA}
\author{Alec Kirkley}
\affiliation{Department of Physics, University of Michigan, Ann Arbor, Michigan 48109, USA}
\author{M. E. J. Newman}
\affiliation{Department of Physics, University of Michigan, Ann Arbor, Michigan 48109, USA}
\affiliation{Center for the Study of Complex Systems, University of Michigan, Ann Arbor, Michigan 48109, USA}

\begin{abstract}
  The friendship paradox is the observation that the degrees of the neighbors of a node in any network will, on average, be greater than the degree of the node itself.  In common parlance, your friends have more friends than you do.  In this paper we develop the mathematical theory of the friendship paradox, both in general as well as for specific model networks, focusing not only on average behavior but also on variation about the average and using generating function methods to calculate full distributions of quantities of interest.  We compare the predictions of our theory with measurements on a large number of real-world network data sets and find remarkably good agreement.  We also develop equivalent theory for the generalized friendship paradox, which compares characteristics of nodes other than degree to those of their neighbors.
\end{abstract}

\maketitle

\section{Introduction}
It may appear to you that your friends are more popular than you are, and if so, you may well be right.  It is however not necessarily your fault.  It is, rather, a natural consequence of network structure.  Feld~\cite{Feld91} has shown that in any network the average degree (i.e.,~the number of neighbors) of the neighbor of a node is strictly greater than the average degree of nodes in the network as a whole.  Applied to networks of friendship, this implies that on average your friends have more friends than you do.  This phenomenon is known as the \emph{friendship paradox}.

A related phenomenon, the \defn{generalized friendship paradox}, describes similar behavior with respect to other attributes of network nodes~\cite{JE14}.  Are your friends richer than you, for instance, or smarter, or more attractive?  Generalized friendship paradoxes arise when such attributes are correlated with node degree.  If richer people are on average also more popular, then wealth and popularity will be positively correlated and hence the tendency for your friends to be more popular than you could mean they are also richer.  Examples of generalized friendship paradoxes occur for instance with citation counts in collaboration networks (your collaborators have more citations than you do)~\cite{EJ14} and viral content dissemination in online social networks (your online friends receive more viral content than you do)~\cite{HKL13}.

Understood as a mathematical statement about network averages, the friendship paradox occurs in all networks.  How the effect is manifested, however, depends on the details of a network's structure.  For any given person we can measure the difference between how popular their friends are and how popular they are.  The results of Feld~\cite{Feld91} tell us that this difference must be positive on average, but to accurately characterize the effect we should consider the entire distribution of differences.  How large will the differences be?  Is their average driven by a handful of outliers?  How much variation is there?  For how many people will the difference be positive?  The answers to such questions all depend on the specific details of the network under study.

In this paper we develop the theory of the friendship paradox in both its original and generalized versions.  We derive expressions for the full distribution of differences in three progressively more complex network models and find that the observed distributions in real-world networks are in good agreement with our theoretical results.

\section{The friendship paradox}
\label{sec:paradox}
Informally, the friendship paradox states that people's friends tend to be more popular than they themselves are.  Stated a little more precisely, nodes in a network tend to have lower degree than their neighbors do.  Consider an undirected network of $n$ nodes, labeled by integers $i=1,\ldots, n$.  For any given node~$i$ we can compute the difference~$\Delta_i$ between the average of its neighbors' degrees and its own degree:
\begin{equation}
\Delta_i = \frac{1}{k_i} \sum_j A_{ij}k_j - k_i,
\label{eq:Delta}
\end{equation}
where $A_{ij}$ is an element of the adjacency matrix and $k_i = \sum_j A_{ij}$ is the degree of node~$i$.  (The value of $\Delta_i$ is undefined for nodes with degree zero, since they have no neighbors.  We will assume that there are no such nodes in the network, or that all of them have been removed before the analysis.)

One statement of the friendship paradox is that the average of $\Delta_i$ across all nodes is greater than zero, which can be proven as follows.  We write the average as
\begin{align}
\frac{1}{n} \sum_{i=1}^n \Delta_i &=
  \frac{1}{n} \sum_i \biggl( \frac{1}{k_i} \sum_j A_{ij}k_j - k_i
  \biggr) \nonumber\\
  &= \frac{1}{n} \sum_{ij} \biggl( A_{ij} \frac{k_j}{k_i} - A_{ij} \biggr)
   = \frac{1}{n} \sum_{ij} A_{ij} \biggl( \frac{k_j}{k_i} - 1 \biggr).
\label{eq:fp1}
\end{align}
Exchanging the summation variables $i$ and~$j$ and adding the result to Eq.~\eqref{eq:fp1}, this result can also be written 
\begin{align}
\frac{1}{n} \sum_{i=1}^n \Delta_i &= \frac{1}{2n} \sum_{ij} A_{ij} \biggl( \frac{k_j}{k_i} + \frac{k_i}{k_j} - 2 \biggr) \nonumber\\
  &= \frac{1}{2n} \sum_{ij} A_{ij} \biggl( \sqrt{\frac{k_j}{\smash[b]{k_i}}} -
  \sqrt{\frac{k_i}{\smash[b]{k_j}}} \biggr)^2 \ge 0.
\label{eq:FP_proof}
\end{align}
The exact equality holds only when $k_i=k_j$ for all pairs of neighboring nodes, i.e.,~when the network is a regular graph (or, technically, when every component is a regular graph).  In all other cases, the average difference between the mean degree of a node's neighbors and its own degree is strictly greater than zero.

For the generalized friendship paradox, which considers attributes other than degree, one can define an analogous quantity~$\Delta_i^{(x)}$ for an attribute~$x$ according to
\begin{equation}
\Delta_i^{(x)} = \frac{1}{k_i} \sum_j A_{ij} x_j - x_i,
\label{eq:Delta_i^x}
\end{equation}
which measures the difference between the average of the attribute for node~$i$'s neighbors and the value for $i$ itself.  When the average of this quantity over all nodes is positive one may say that the generalized friendship paradox holds.  In contrast to the case of degree, this is not always true---the value of~$\Delta_i^{(x)}$ can be zero or negative---but we can write the average as
\begin{align}
\frac{1}{n} \sum_i \Delta_i^{(x)}
  &= \frac{1}{n} \sum_i \biggl( \frac{1}{k_i} \sum_j A_{ij} x_j - x_i \biggr)
     \nonumber\\
 &= \frac{1}{n} \sum_i \biggl( x_i \sum_j \frac{A_{ij}}{k_j} - x_i \biggr),
\end{align}
where the second line again follows from interchanging summation indices.  Defining the new quantity
\begin{equation}
\kappa_i = \sum_j \frac{A_{ij}}{k_j}
\end{equation}
and noting that
\begin{equation}
\frac{1}{n} \sum_i \kappa_i = \frac{1}{n} \sum_{ij} \frac{A_{ij}}{k_j}
  = \frac{1}{n} \sum_j {1\over k_j} \sum_i A_{ij} = 1,
\end{equation}
we can then write
\begin{align}
\frac{1}{n} \sum_i \Delta_i^{(x)}
  &= \frac{1}{n} \sum_i x_i\kappa_i - \frac{1}{n} \sum_i x_i \>
     \frac{1}{n} \sum_i \kappa_i \nonumber\\
  &= \cov(x,\kappa).
\label{eq:covariance}
\end{align}
Thus we will have a generalized friendship paradox in the sense defined here if (and only if) $x$ and $\kappa$ are positively correlated.

This is perhaps not exactly the result one might have anticipated when, in the introduction, we argued that generalized friendship paradoxes arise when an attribute~$x$ is positively correlated with degree---it turns out that correlation with the more complex quantity~$\kappa$ is the crucial behavior.  Degree, however, is always positively correlated with~$\kappa$ since, combining Eqs.~\eqref{eq:FP_proof} and~\eqref{eq:covariance}, we have $\cov(k,\kappa) = (1/n) \sum_i \Delta_i \ge 0$.  While it is not mathematically guaranteed, in practice we therefore expect properties that correlate with degree to also correlate with~$\kappa$ and hence in such situations we can reasonably expect to observe a generalized friendship paradox.

\section{Models}
While these results are illuminating, there is more to be said on this topic.  We have shown that the network average of $\Delta_i$ will always be greater than zero, but we have not said by how much or what the value depends on.  Nor have we considered how $\Delta_i$ varies across a network.  As an example (albeit a somewhat contrived one), consider a 1000-node network consisting of a complete graph with one edge removed.  In such a network almost all of the nodes---998 of them---have a degree \emph{larger} than the average of their neighbors, but there are two outliers that substantially skew the distribution so that over the whole network nodes are still less popular than their neighbors on average.  In this case, therefore, the average is not very informative.  To understand the friendship paradox fully we need to move beyond statements about averages.

In doing so, however, the specific structure of the network becomes important.  A common way to study the effects of structure is to examine the behavior of model networks and this is the approach we take in the remainder of this paper, considering the friendship paradox for three successively more complex network models, the Poisson random graph (sometimes called the Erd\H{o}s--R\'enyi model)~\cite{gilbert_random_1959,ER60,Bollobas01}, the configuration model~\cite{bollobas_probabilistic_1980,NSW01}, and a more sophisticated model that incorporates degree correlations.

\subsection{Poisson random graph}
\label{sec:rg}
The Poisson random graph is the simplest of random graph models and one of the most widely studied.  In this model one takes $n$ nodes and connects each pair independently with some probability~$p$.  When the network is large and sufficiently sparse---when $n\to \infty$ and $p=\lambda/(n-1)$ with $\lambda$ growing slower than~$n$ so that $p\to0$---the degrees are Poisson distributed with mean~$\lambda$, meaning that the probability~$p_k$ of a node having degree~$k$ is
\begin{equation}
p_k = {\lambda^k\over k!} \, e^{-\lambda}.
\label{eq:poisson}
\end{equation}

Let us compute the complete distribution within this model of the quantity~$\Delta_i$ defined in Eq.~\eqref{eq:Delta}.  To compute this distribution, we note that
\begin{equation}
P(\Delta) = \sum_k p_k P(\Delta|k),
\label{eq:Poisson_P_delta0}
\end{equation}
where $P(\Delta|k)$ denotes the probability that a node has value~$\Delta$ given that it has degree~$k$.  For given $\Delta$ and~$k$, Eq.~\eqref{eq:Delta} tells us that the sum of the neighboring degrees of the node is $\sum_j A_{ij} k_j = k \Delta + k^2$.  Let us denote this sum by~$K$.  Then we have
\begin{equation}
P(\Delta|k) = P( K = k \Delta +k^2 |k).
\label{eq:Poisson_P_delta}
\end{equation}
We know that $K$~is an integer with $K\ge k$, since each of the node's $k$ neighbors necessarily has degree at least~1.  Hence the allowed values of $\Delta$ that satisfy $K = k \Delta +k^2$ must be rational numbers of the form $\Delta = 1-k+m/k$ where $m$ is a non-negative integer.

To evaluate~\eqref{eq:Poisson_P_delta} we need to know the distribution of the sum~$K = \sum_j A_{ij} k_j$ of neighbor degrees.  While nodes in general follow the degree distribution~$p_k$, the degrees of neighbors follow a modified distribution.  A neighbor, by definition, is a node arrived at by following an edge and each node of degree~$k$ is at the end of $k$ edges, so the degree distribution for nodes at the ends of edges goes not as $p_k$ but as $kp_k$, which after appropriate normalization gives a probability distribution~$q_k$ for neighbor degrees of the form
\begin{equation}
q_k = \frac{k p_k}{\sum_j j p_j}.
\label{eq:qk}
\end{equation}
For the Poisson degree distribution of Eq.~\eqref{eq:poisson}, we then find that
\begin{equation}
q_k = {k \lambda^k e^{-\lambda}\over \lambda k!}
    = {\lambda^{k-1}\over (k-1)!} e^{-\lambda}.
\end{equation}
In other words $q_k$~is also a Poisson distribution, but shifted by one, meaning that $k-1$ is a Poisson variable with mean~$\lambda$.

We can write the sum~$K$ as
\begin{equation}
K = \sum_j A_{ij} k_j = k_i + \sum_j A_{ij} (k_j-1).
\end{equation}
But $\sum_j A_{ij} (k_j-1) \sim \text{Poisson}(k_i\lambda)$, since the sum of the Poisson variables~$(k_j-1)$ is itself a Poisson random variable with mean a factor of~$k_i$ greater.  Hence
\begin{equation}
P(\Delta|k) = P(K = k \Delta + k^2 | k)
	    = {(k\lambda)^{k \Delta +k^2-k}\over(k \Delta +k^2-k)!} e^{-k\lambda}.
\end{equation}
For each value of~$k$, this gives us a distribution over a discrete set of equally spaced values $\Delta=1-k+m/k$ with $m=0\ldots\infty$, and the full distribution, Eq.~\eqref{eq:Poisson_P_delta0}, is a linear combination of an infinite number of such distributions.

\begin{figure}
\includegraphics[width=1\linewidth]{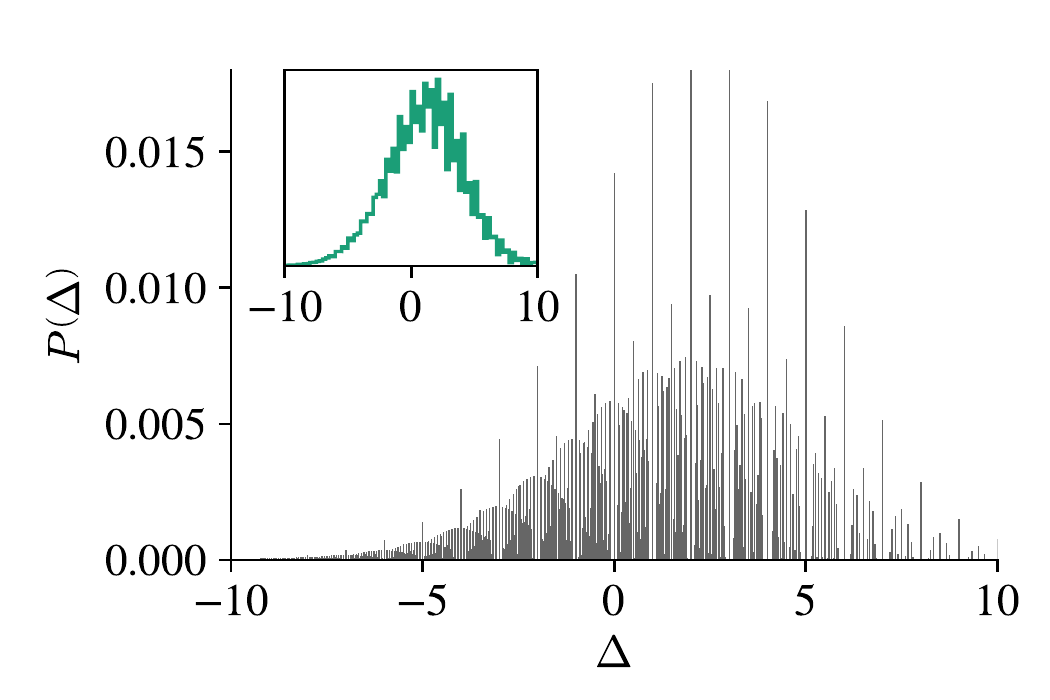}
\caption{The distribution of the quantity~$\Delta$, defined in Eq.~\eqref{eq:Delta}, for a random graph with mean degree~8.  The main panel shows the probability mass function while the inset shows a traditional histogram of the distribution.}
\label{fig:poisson_delta}
\end{figure}

Figure~\ref{fig:poisson_delta} shows an example of the resulting distribution of~$\Delta$ computed from Eq.~\eqref{eq:Poisson_P_delta0} for the case $\lambda=8$.  The individual discrete distributions are clearly visible as sets of equally spaced peaks and the overall resulting distribution is quite complicated---both widely dispersed and jagged, even for this simple network model.  Note also that a significant fraction of nodes have~$\Delta<0$, meaning that they do not satisfy the traditional definition of the friendship paradox---they have more friends than their average neighbor does.

It is also possible to calculate the mean and variance of~$\Delta$ for the Poisson random graph.  We find that
\begin{equation}
E[\Delta] = 1; \quad \var(\Delta) = \lambda (1+ E[k^{-1}]).
\end{equation}
(Note that again we remove any nodes of degree zero before computing~$E[k^{-1}]$.)  While the expected value of $\Delta$ is always 1, the variance is larger than the mean degree~$\lambda$, meaning that more and more nodes have $\Delta<0$ as $\lambda$ becomes large, with the fraction tending to a half.  For example, in Fig.~\ref{fig:poisson_delta}, where the mean degree is 8, around $35\%$ of nodes have degree larger than that of their average neighbor.  This rises to $44\%$ when the mean degree is~64, and $49\%$ for a mean degree of 1024.  For large~$\lambda$ therefore, no meaningful ``friendship paradox'' applies.  Here, as is often the case, looking only at the average value of the distribution is misleading when there is large variation.

\subsection{The configuration model}
\label{sec:cm}
The random graph of the previous section is in many respects not a realistic model.  In particular, as we have noted, it has a Poisson degree distribution, which is very different from the broad degree distributions seen in typical real-world networks~\cite{BA99b,ASBS00}.  We can address this shortcoming by using a more sophisticated random graph model that allows for arbitrary degree distributions, the so-called configuration model~\cite{bollobas_probabilistic_1980,NSW01}.  In this model one fixes the degree of each of the nodes and then draws a network at random from the set of all networks with the given degrees.

Calculations on networks such as the configuration model can be greatly simplified by using generating function methods~\cite{Wilf94,Newman18c}.  For instance, many properties of the model can be expressed in terms of the generating function for the degree distribution~$p_k$, defined by
\begin{equation}
f(z) = \sum_{k=0}^{\infty} p_k z^k.
\label{eq:gen_fun_defn}
\end{equation}
We employ this approach here too, but there is a catch, in that generating functions are normally applied to distributions over integer quantities, like the degree, but the quantity~$\Delta$, whose distribution is our main focus here, can take non-integer values.  To allow for this, we make use of the (two-sided) Laplace transform
\begin{equation}
F(s) = \int_{-\infty}^{\infty} p(x)\,e^{-s x}\>dx,
\label{eq:Laplace_defn}
\end{equation}
which is the standard extension of the generating function to a variable~$x$ on the real line.

However, the distribution of~$\Delta$ is not continuous-valued either.  It is nonzero on a dense set of rational values but zero everywhere else---see Fig.~\ref{fig:poisson_delta}.  To allow for this, we consider functions~$p(x)$ that are equal to a discrete sum of Dirac delta functions.  For the degree distribution~$p_k$, for instance, we would define
\begin{equation}
p(x) = \sum_{k=0}^{\infty} p_k \delta(x-k).
\end{equation}
With this definition Eqs.~\eqref{eq:gen_fun_defn} and~\eqref{eq:Laplace_defn} are essentially equivalent, since
\begin{align}
F(s) &= \int_{-\infty}^{\infty} \sum_k p_k \delta(x-k)\,e^{-s x}\>dx
      = \sum_k p_k e^{-s k} \nonumber\\
     &= f(e^{-s}),
\end{align}
but Eq.~\eqref{eq:Laplace_defn} also allows for quantities like~$\Delta$ that have non-integer values.

The Laplace transform has several properties that will be useful for our purposes.  First, if $x$ is a random variable whose distribution has Laplace transform~$F_x(s)$, then the Laplace transform for the distribution of $ax+b$~is
\begin{equation}
F_{ax+b}(s) = e^{-b s} f_x(a s).
\label{eq:Laplace_linear}
\end{equation}
Second, if $x$ and $y$ are independent random variables, then the Laplace transform for their sum is the product of the Laplace transforms for $x$ and $y$ alone:
\begin{equation}
F_{x+y}(s) = F_x(s) F_y(s).
\label{eq:Laplace_sum}
\end{equation}
These two results now allow us to calculate~$F_\Delta(s)$, the Laplace transform for~$P(\Delta)$.  We follow essentially the same logic as we did for the Poisson random graph: we consider the Laplace transform for the distribution of~$\Delta$ for fixed degree, then we average over degree.

If node~$i$ has degree~$k_i$, then from Eq.~\eqref{eq:Delta} $\Delta_i$ is
\begin{equation}
\Delta_i = \sum_j A_{ij} \biggl( {k_j \over k_i} - 1 \biggr).
\label{eq:Delta_cond_k_sum}
\end{equation}
Each $k_j$ is the degree of a neighbor node which, as before, is a random quantity distributed according to $q_k \propto k p_k$.  Let $G(s)$ be the Laplace transform for this distribution:
\begin{equation}
G(s) = \sum_{k=0}^{\infty} q_k e^{-s k}.
\end{equation}
Then from Eq.~\eqref{eq:Laplace_linear} the Laplace transform for $k_j/k_i-1$ is $e^s\,G(s/k_i)$ and, since each of the $k_i$ nonzero terms in the sum of Eq.~\eqref{eq:Delta_cond_k_sum} is independent, the Laplace transform for the full sum is $[ e^{s} G(s/k_i) ]^{k_i}$ by Eq.~\eqref{eq:Laplace_sum}.

This is for a node of degree~$k_i$.  Since the Laplace transform is linear, we can now simply average over degree to compute the transform for the full distribution of~$\Delta$:
\begin{align}
F_\Delta(s) = \sum_k p_k\,e^{s k}\,G(s/k)^k.
\label{eq:F_Delta_config}
\end{align}
Given any degree distribution $p_k$ we can use this equation to compute~$F_{\Delta}(s)$.  Inverting the Laplace transform then gives us the density function~$\rho(x)$ of $\Delta$ itself:
\begin{align}
\rho(x) &= \sum_{\Delta} P(\Delta) \delta(x - \Delta) \nonumber\\
 &= {1 \over 2 \pi} \int_{-\infty}^{+\infty} F_\Delta(i s)\,e^{i s x}\>ds,
\label{eq:F_delta_inverse}
\end{align}
where the sum over $\Delta$ is a sum over all rational numbers---all possible values of $\Delta$.

Since $P(\Delta)$ is a rather complicated object, it is in practice simpler to integrate $\rho(x)$ to compute the probability that $\Delta$ falls between any two values---in other words a histogram of~$\Delta$.  In fact, we can do something more sophisticated: we can use any kernel we like to calculate a kernel density estimate of the distribution of~$\Delta$.  For a general kernel function~$\kappa(x)$ with Laplace transform~$F_\kappa(s)$ we have
\begin{align}
\rho_\kappa(x) &= \sum_{\Delta} P(\Delta) \kappa(x - \Delta) \nonumber\\
	&= {1 \over 2 \pi} \int_{-\infty}^{+\infty} F_\Delta(i s) F_\kappa(i s)
     \,e^{i s x}\>ds.
\label{eq:kde_integral}
\end{align}
A conventional histogram is equivalent to using a rectangular (``top hat'') kernel, but for our figures we use a smoother double-exponential kernel (also known as a Laplace distribution):
\begin{equation}
\kappa(x) = {1 \over 2b} e^{- \vert x \vert /b },
\end{equation}
where the parameter~$b$ sets the width of the distribution.  (We arbitrarily pick $b=1/3$.)  The Laplace transform for this choice of $\kappa(x)$ is
\begin{equation}
F_\kappa(s) = { 1 \over 1 - b^2 s^2 }.
\end{equation}

\begin{figure}
\includegraphics[width=0.9\linewidth]{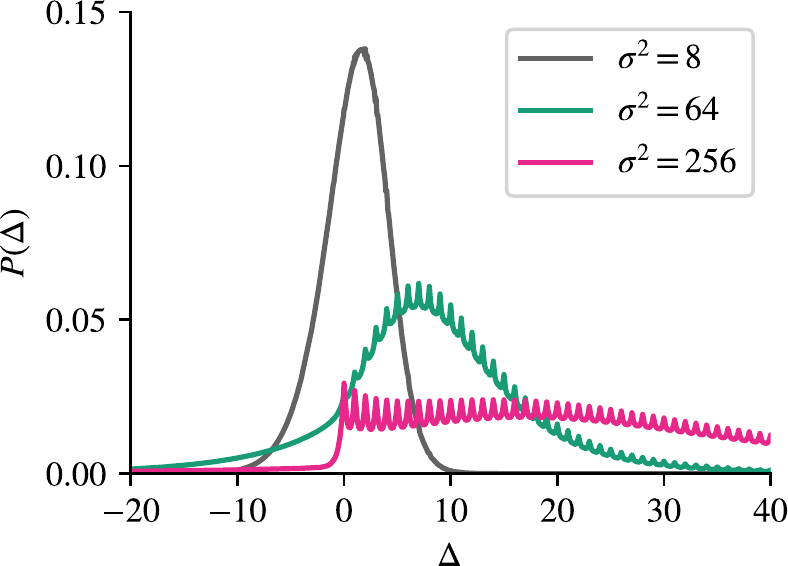}
	\caption{The distribution of $\Delta_i$ for three configuration models with the truncated power-law degree distribution \mbox{$p_k \propto k^{-\alpha} e^{-\beta k}$} and different choices of the parameters~$\alpha$ and~$\beta$.  $\Delta_i$~can take only rational values, but for clarity we show a kernel density estimate of the distribution, calculated from Eq.~\eqref{eq:kde_integral} with a Laplace distribution kernel.  The values of $\alpha$ and $\beta$ were chosen so that the mean degree is always~$8$, while the variance is $8$, $64$, or $256$ as indicated.}
\label{fig:P_delta_var}
\end{figure}

Figure~\ref{fig:P_delta_var} shows the distribution of~$\Delta$ computed in this way for configuration models with the truncated power-law degree distribution
\begin{equation}
p_k \propto k^{-\alpha} e^{-\beta k}
\label{eq:power_law_with_cutoff}
\end{equation}
and three different choices of the parameters $\alpha$ and~$\beta$.  Each example has the same mean degree but, as the figure shows, the distributions of~$\Delta$ are quite different.

\subsection{Random graphs with degree correlations}
\label{sec:assortative}
The configuration model of the previous section improves on the Poisson random graph by allowing arbitrary distributions of node degrees, but like the random graph it lacks any correlation or assortativity between the degrees of adjacent nodes.  Such correlation is common in real-world networks \cite{Newman02f,NP03b,HuWang09} and will clearly impact friendship paradox phenomena.

One can create a model network with degree correlations by fixing not only the degree distribution~$p_k$ as in the configuration model, but the joint distribution of adjacent degrees~$Q_{jk}$, which is the fraction of edges that join nodes of degrees $j$ and~$k$~\cite{Newman02f}.  Note that the distribution of degrees at the end of an edge is then given by $q_k = \sum_j Q_{j k}$, so that fixing $Q_{j k}$ also fixes the degree distribution.

The calculation of the distribution of~$\Delta$ proceeds in a similar manner to that for the configuration model.  If we follow an edge that begins at a node of degree~$k$, it will end up at a node of degree~$j$ with probability $Q_{j k}/q_k$, meaning that the Laplace transform for the degrees of neighbors is
\begin{equation}
	G_k(s) = \sum_j \frac{Q_{j k}}{q_k} e^{-s j}.
\end{equation}
Note that this function depends on the degree $k$ of the node at which we started.  Nevertheless, the calculation proceeds essentially as before.  The equivalent of Eq.~\eqref{eq:F_Delta_config}~is
\begin{equation}
F_{\Delta}(s) = \sum_{k} p_k\,e^{s k}\,G_k(s/k)^k,
\label{eq:F_Delta_cor_config}
\end{equation}
and the distribution of~$\Delta$ can be calculated from $F_{\Delta}(s)$ using Eq.~\eqref{eq:kde_integral}.

Degree correlations are commonly quantified using an assortativity coefficient~$r$, defined as the Pearson correlation coefficient of degrees across edges~\cite{Newman02f}.  In terms of the quantities defined here,
\begin{equation}
r = \frac{\sum_{jk} j k ( Q_{jk} - q_j q_k)}{\sigma_{q}^2},
\end{equation}
where $\sigma_{q}^2$ is the variance of the distribution $q_k$.  To study how the effects of the friendship paradox vary with varying~$r$ it is convenient to define a model network that allows us to adjust~$r$, in effect a correlated version of the configuration model, parameterized by its degree distribution and a single extra parameter controlling the assortativity.  This means choosing a suitable value of~$Q_{jk}$, which we do by maximizing the entropy
\begin{equation}
S(Q) = -\sum_{jk} Q_{jk} \log Q_{j k}
\end{equation}
subject to the constraints $\sum_j Q_{jk} = q_k = kp_k/\sum_j jp_j$ for all~$k$ and $\sum_{jk} j k ( Q_{jk} - q_j q_k)/ \sigma_{q}^2 = r$.  The maximum entropy distribution is often considered to be the least biased choice for a given set of constraints, meaning that it makes no assumptions other than those implied by the constraints themselves.

The maximum entropy solution for~$Q_{jk}$ in this case is
\begin{equation}
Q_{jk} = \frac{e^{\gamma j k}}{Z_j Z_k} \, q_j q_k\,,
	\label{eq:max_ent_Q}
\end{equation}
where $\gamma$ is a Lagrange multiplier whose value controls the assortativity and the $Z_k$ are normalizing constants that satisfy the equations
\begin{equation}
Z_k = \sum_j \frac{q_j e^{\gamma j k}}{Z_j},
\end{equation}
which we solve numerically by iteration.  This model defines an ensemble of random networks with a desired level of assortativity and allows us to study the generic effects of assortativity on network properties, including the friendship paradox.

\begin{figure}
\includegraphics[width=0.9\linewidth]{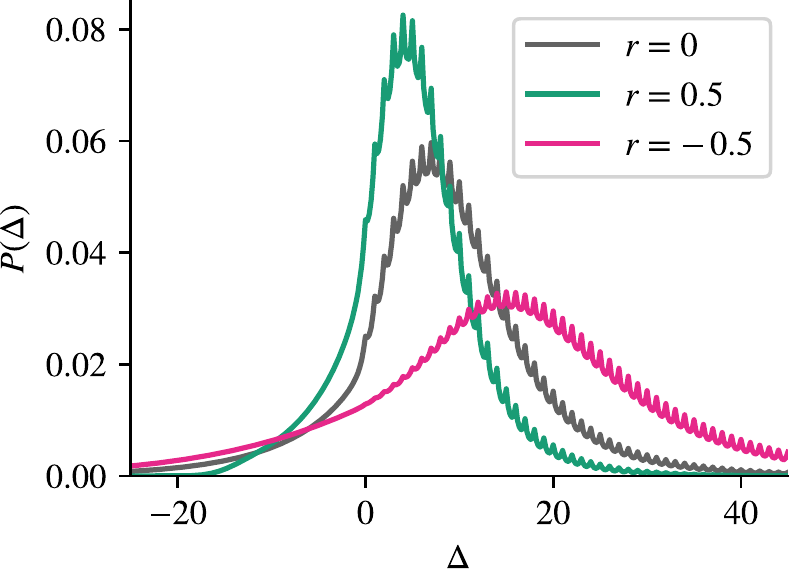}
\caption{\label{fig:P_delta_assortativity}Distribution for $\Delta$ in a random graphs with $p_k$ defined by Eq.~\eqref{eq:power_law_with_cutoff}. The mean degree $\left< k \right>=8$ and $\sigma_k^2 = 64$. Three different values of $\gamma$ were chosen so that the assortativity coefficient $r$ takes the values $0$, $0.5$, and $-0.5$.}
\end{figure}

Figure~\ref{fig:P_delta_assortativity}, for example, shows the probability distribution of~$\Delta$ for fixed degree distribution and three different choices of the assortativity coefficient~$r$.  Note how the average value of $\Delta$ decreases as the networks become more assortative, but so too does the variance, leading to a more complex picture.  This is a good example of why we should be wary of conclusions based on the average value of~$\Delta$ alone.

\begin{figure}
\begin{center}
\includegraphics[width=0.9\linewidth]{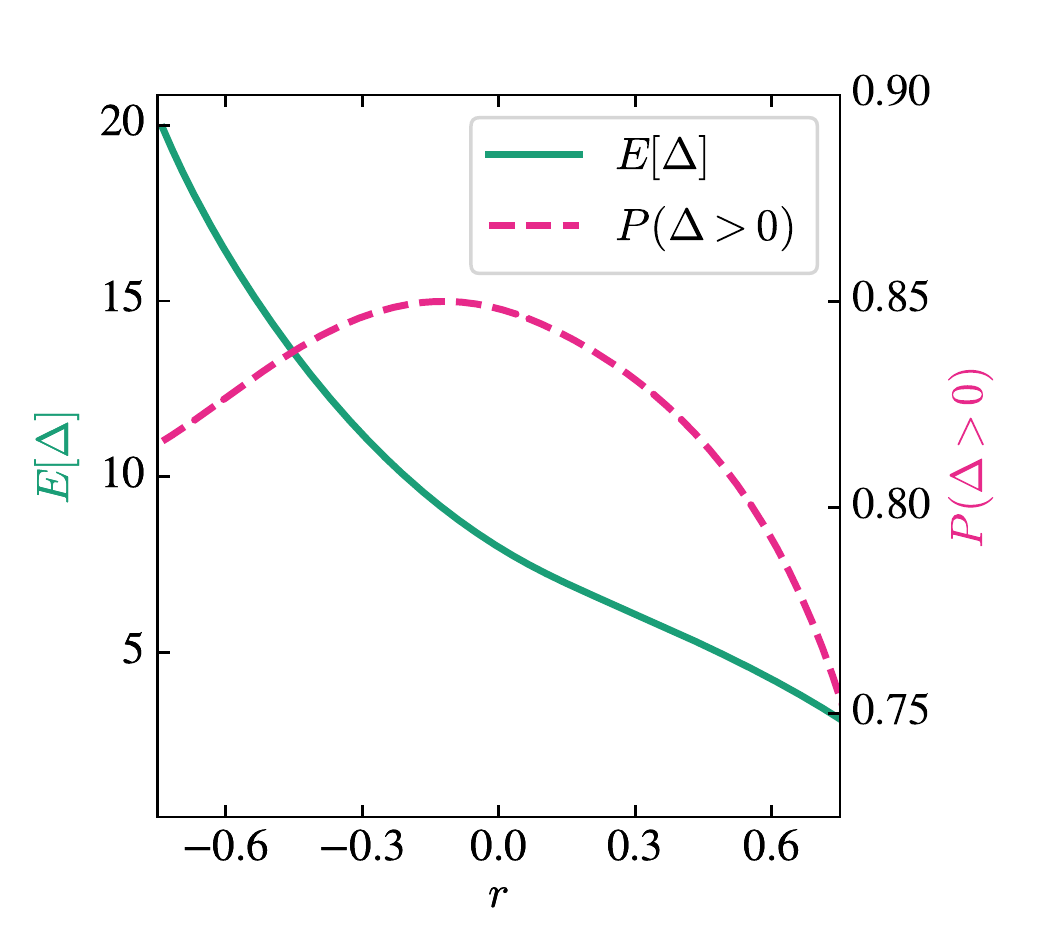}
\end{center}
\caption{Two measures of the magnitude of the friendship paradox as a function of assortativity. The degree distribution is that of Eq.~\eqref{eq:power_law_with_cutoff} with $\alpha$ and $\beta$ chosen so that $\left< k \right>=8$ and $\sigma_k^2 = 64$.}
\label{fig:E_delta}
\end{figure}

Figure~\ref{fig:E_delta} sheds more light on this point, showing the average value along with the expected fraction \mbox{$P(\Delta>0)$} of nodes with positive~$\Delta$, both as a function of~$r$.  Both of these quantities can be viewed as measures of the strength of the friendship paradox, but they behave in different ways.  While $E[\Delta]$ does indeed decrease monotonically with assortativity, the fraction of nodes with positive~$\Delta$---in effect, the fraction of nodes that display the classic friendship paradox behavior---peaks at small negative~$r$, and has lower values for both large positive and large negative~$r$.

\subsection{Comparison with real-world networks}
\begin{figure}
\begin{center}
\includegraphics[width=0.775\linewidth]{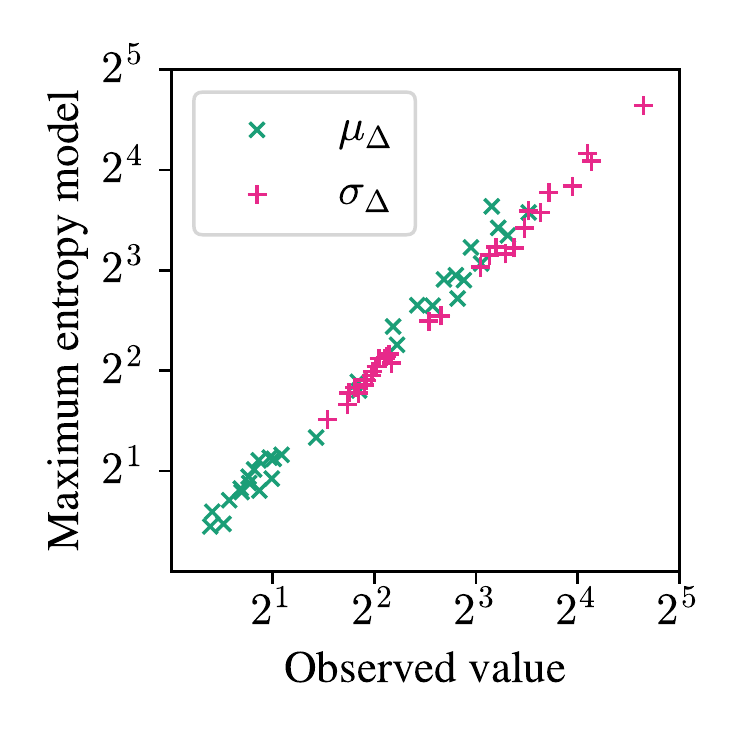}
\end{center}
	\caption{Mean and standard deviation of $\Delta$ in 32 real-world social networks \cite{GD03,BPDA04b,Newman06c,Newman01a,GDDGA03,harris_national_2008,MB17,Knuth93,BS16,Weeks2002,Lusseau03a,FKESR12,Tang2009,Edwards15,SEIERSTAD201144,MFB15} compared to predictions from the maximum-entropy model, Eq.~\eqref{eq:max_ent_Q}.
	}
\label{fig:realworld}
\end{figure}

How good a guide are these model calculations to the behavior of real-world networks?  To shed light on this question we compare the mean and variance of $\Delta$ in 32 real-world social networks with values calculated from the assortative network model of the previous section with the same~$r$.  (The mean and variance in the model can be computed from the first and second derivatives of Eq.~\eqref{eq:F_Delta_cor_config}.)  The results are shown in Fig.~\ref{fig:realworld} and, as we can see, there is remarkably good agreement between theoretical and empirical results---we find $R^2$ values of $0.93$ and $0.99$ between theory and experiment for the mean and standard deviation of~$\Delta$ respectively.  By comparison, the standard configuration model, which fixes the degree distribution only, gives $R^2$ values of $0.77$ and~$0.95$.  Thus it would be fair to say that the distribution of~$\Delta$ is fairly accurately captured by the degree distribution alone, but that the inclusion of assortativity results in a significant improvement.

\subsection{Generalized friendship paradox}
Before finishing let us return to the generalized friendship paradox.  Recall that for any quantity~$x$ defined on the nodes of a network the quantity~$\Delta_{i}^{(x)}$, Eq.~\eqref{eq:Delta_i^x}, measures the difference between the average value of $x$ at node $i$'s neighbors and $i$'s own value.  We can compute the distribution of~$\Delta^{(x)}$, for instance for the degree-correlated model of Section~\ref{sec:assortative}, using the Laplace transform formalism again.  The argument differs from previous developments in some details but remains conceptually similar.  One first writes the Laplace transform for the distribution of a node's value of $x$ given its degree~$k$
\begin{equation}
H_k(s) = \int_{-\infty}^\infty P(x \vert k)\,e^{-s x}\>dx,
\end{equation}
where $P(x \vert k)$ is the probability that a node of degree $k$ has value~$x$.  Since the neighbors of a degree-$k$ node have degree distributed as $Q_{j k} / q_{k}$, they have a distribution of $x$ values with Laplace transform
\begin{equation}
	G_k(s) = \sum_j { Q_{k j} \over q_k } H_{j}(s).
\end{equation}
Applying the key properties in Eqs.~\eqref{eq:Laplace_linear} and~\eqref{eq:Laplace_sum}, we then arrive at
\begin{equation}
F_{\Delta^{(x)}}(s) = \sum_k p_k G_{k}(s/k)^{k} H_k(-s),
\label{eq:gfp_F_Delta}
\end{equation}
and inverting $F_{\Delta^{(x)}}$ gives the distribution for~$\Delta^{(x)}$.

As an example, we have tested Eq.~\eqref{eq:gfp_F_Delta} for normally distributed~$x$ linearly correlated with degree and find behavior closely similar to that of Fig.~\ref{fig:P_delta_assortativity}.  Behavior like this could have an impact in any situation where the generalized friendship paradox has practical consequences.  For example, it has been found that people's individual well-being can be substantially affected by the behavior of their network neighbors.  People whose acquaintances smoke are more likely to smoke themselves~\cite{CF08}.  People whose friends appear to be better off than they are may develop a lower sense of self-worth~\cite{kross2013facebook,LYW16}.  If people with more acquaintances tend to smoke more, or if well-off people with exciting lives have a lot of friends or followers on social media, then we may have a generalized friendship paradox in which you are most likely to have contact with precisely those people who would adversely affect you.

As another example, it has been shown that polling forecasts of election outcomes can be significantly improved by focusing not on how study participants say they will vote but on how they expect their acquaintances to vote~\cite{nettasinghe2019your}, in part because this reduces variance in the estimates of outcomes.  If voting intention is subject to the generalized friendship paradox however---if for instance partisan inclination is correlated with degree---then the tendency for one's friends to have high degree will cause the resulting sample of the population to be biased and introduce systematic errors~\cite{rothschild2009forecasting}.  The formalism developed here allows us to quantify these effects not only in terms of the average individual but in terms of the complete distribution of outcomes over the entire population.

\section{Conclusions}
In this paper we have quantified the friendship and generalized friendship paradoxes in terms of the difference~$\Delta$ between the characteristics of a node and the average of the same characteristics for the node's neighbors.  Previous studies have examined the mean of this difference but, as we have argued here, to get a full picture one must examine the complete distribution of values.  We have performed theoretical calculations of this distribution for three classes of model networks, the Poisson random graph, the configuration model, and a model of a random degree-assortative network.  Among other things, our results indicate that the friendship paradox will tend to be strongest in networks with very heterogeneous degree distributions and negative assortativity.  Conversely, the effects will tend to be muted when degrees are fairly homogeneous and the network is degree assortative.  On the other hand, we have also seen that even in simple network models the distribution for $\Delta$ can be widely dispersed, meaning that the average value offers an incomplete description of the behavior.

We have also compared our results with a selection of real-world networks, finding remarkably good agreement between theoretical predictions and empirical measurements, particularly in the case of the model that incorporates assortativity.

\begin{acknowledgments}
This work was funded in part by the US Department of Defense NDSEG fellowship program (AK) and by the US National Science Foundation under grants DMS--1710848 and DMS--2005899 (MEJN).

This research uses data from Add Health, a program project directed by Kathleen Mullan Harris and designed by J. Richard Udry, Peter S. Bearman, and Kathleen Mullan Harris at the University of North Carolina at Chapel Hill, and funded by grant P01--HD31921 from the Eunice Kennedy Shriver National Institute of Child Health and Human Development, with cooperative funding from 23 other federal agencies and foundations. Information on how to obtain the Add Health data files is available on the Add Health website (https://addhealth.cpc.unc.edu/). No direct support was received from grant P01-HD31921 for this analysis.
\end{acknowledgments}


\begin{thebibliography}{10}
\expandafter\ifx\csname url\endcsname\relax
  \def\url#1{\texttt{#1}}\fi
\expandafter\ifx\csname urlprefix\endcsname\relax\def\urlprefix{URL }\fi

\bibitem{Feld91}
S.~Feld, Why your friends have more friends than you do. \textit{Am. J.
  Sociol.} \textbf{96}, 1464--1477 (1991).

\bibitem{JE14}
H.-H. Jo and Y.-H. Eom, Generalized friendship paradox in networks with tunable
  degree-attribute correlation. \textit{Phys. Rev. E} \textbf{90}, 022809
  (2014).

\bibitem{EJ14}
Y.-H. Eom and H.-H. Jo, Generalized friendship paradox in complex networks: The
  case of scientific collaboration. \textit{Scientific Reports} \textbf{4},
  4603 (2014).

\bibitem{HKL13}
N.~O. Hodas, F.~Kooti, and K.~Lerman, Friendship paradox redux: Your friends
  are more interesting than you. In \textit{Proceedings of the 7th
  International AAAI Conference on Weblogs and Social Media}, AAAI Press, Palo
  Alto, CA (2013).

\bibitem{gilbert_random_1959}
E.~N. Gilbert, Random graphs. \textit{Annals of Mathematical Statistics}
  \textbf{30}, 1141--1144 (1959).

\bibitem{ER60}
P.~Erd\H{o}s and A.~R\'enyi, On the evolution of random graphs.
  \textit{Publications of the Mathematical Institute of the Hungarian Academy
  of Sciences} \textbf{5}, 17--61 (1960).

\bibitem{Bollobas01}
B.~Bollob\'as, \textit{Random Graphs}. Academic Press, New York, 2nd edition
  (2001).

\bibitem{bollobas_probabilistic_1980}
B.~Bollob{\'a}s, A probabilistic proof of an asymptotic formula for the number
  of labelled regular graphs. \textit{European Journal of Combinatorics}
  \textbf{1}, 311--316 (1980).

\bibitem{NSW01}
M.~E.~J. Newman, S.~H. Strogatz, and D.~J. Watts, Random graphs with arbitrary
  degree distributions and their applications. \textit{Phys. Rev. E}
  \textbf{64}, 026118 (2001).

\bibitem{BA99b}
A.-L. Barab\'asi and R.~Albert, Emergence of scaling in random networks.
  \textit{Science} \textbf{286}, 509--512 (1999).

\bibitem{ASBS00}
L.~A.~N. Amaral, A.~Scala, M.~Barth\'elemy, and H.~E. Stanley, Classes of
  small-world networks. \textit{Proc. Natl. Acad. Sci. USA} \textbf{97},
  11149--11152 (2000).

\bibitem{Wilf94}
H.~Wilf, \textit{Generatingfunctionology}. Academic Press, London, 2nd edition
  (1994).

\bibitem{Newman18c}
M.~Newman, \textit{Networks}. Oxford University Press, Oxford, 2nd edition
  (2018).

\bibitem{Newman02f}
M.~E.~J. Newman, Assortative mixing in networks. \textit{Phys. Rev. Lett.}
  \textbf{89}, 208701 (2002).

\bibitem{NP03b}
M.~E.~J. Newman and J.~Park, Why social networks are different from other types
  of networks. \textit{Phys. Rev. E} \textbf{68}, 036122 (2003).

\bibitem{HuWang09}
H.-B. Hu and X.-F. Wang, Disassortative mixing in online social networks.
  \textit{Europhys. Lett.} \textbf{86}, 18003 (2009).

\bibitem{GD03}
P.~Gleiser and L.~Danon, Community structure in jazz. \textit{Advances in
  Complex Systems} \textbf{6}, 565--573 (2003).

\bibitem{BPDA04b}
M.~Bogu{\~n}\'a, R.~Pastor-Satorras, A.~D{\'\i}az-Guilera, and A.~Arenas,
  Models of social networks based on social distance attachment. \textit{Phys.
  Rev. E} \textbf{70}, 056122 (2004).

\bibitem{Newman06c}
M.~E.~J. Newman, Finding community structure in networks using the eigenvectors
  of matrices. \textit{Phys. Rev. E} \textbf{74}, 036104 (2006).

\bibitem{Newman01a}
M.~E.~J. Newman, The structure of scientific collaboration networks.
  \textit{Proc. Natl. Acad. Sci. USA} \textbf{98}, 404--409 (2001).

\bibitem{GDDGA03}
R.~Guimer\`a, L.~Danon, A.~D\'{\i}az-Guilera, F.~Giralt, and A.~Arenas,
  Self-similar community structure in a network of human interactions.
  \textit{Phys. Rev. E} \textbf{68}, 065103 (2003).

\bibitem{harris_national_2008}
K.~M. Harris and J.~R. Udry, National Longitudinal Study of Adolescent to
  Adult Health (Add Health), 1994--2008, Public Use Version 21 (2008).

\bibitem{MB17}
B.~F. Maier and D.~Brockmann, Cover time for random walks on arbitrary complex
  networks. \textit{Phys. Rev. E} \textbf{96}, 042307 (2017).

\bibitem{Knuth93}
D.~E. Knuth, \textit{The Stanford GraphBase: A Platform for Combinatorial
  Computing}. Addison-Wesley, Reading, MA (1993).

\bibitem{BS16}
A.~Beveridge and J.~Shan, Network of thrones. \textit{Math Horizons}
  \textbf{23}(4), 18--22 (2016).

\bibitem{Weeks2002}
M.~R. Weeks, S.~Clair, S.~P. Borgatti, K.~Radda, and J.~J. Schensul, Social
  networks of drug users in high-risk sites: Finding the connections.
  \textit{AIDS and Behavior} \textbf{6}, 193--206 (2002).

\bibitem{Lusseau03a}
D.~Lusseau, K.~Schneider, O.~J. Boisseau, P.~Haase, E.~Slooten, and S.~M.
  Dawson, The bottlenose dolphin community of {D}oubtful {S}ound features a
  large proportion of long-lasting associations. {C}an geographic isolation
  explain this unique trait? \textit{Behavioral Ecology and Sociobiology}
  \textbf{54}, 396--405 (2003).

\bibitem{FKESR12}
M.~Fire, G.~Katz, Y.~Elovici, B.~Shapira, and L.~Rokach, Predicting student
  exam's scores by analyzing social network data. In R.~Huang, A.~A. Ghorbani,
  G.~Pasi, T.~Yamaguchi, N.~Y. Yen, and B.~Jin (eds.), \textit{Active Media
  Technology}, pp. 584--595, Springer, Berlin (2012).

\bibitem{Tang2009}
J.~Tang, J.~Sun, C.~Wang, and Z.~Yang, Social influence analysis in large-scale
  networks. In \textit{Proceedings of the 15th ACM SIGKDD International
  Conference on Knowledge Discovery and Data Mining}, KDD '09, pp. 807--816,
  ACM, New York, NY (2009).

\bibitem{Edwards15}
G.~Edwards, FIFA networks (2015),
\urlprefix\url{https://sites.google.com/site/ucinetsoftware/datasets/covert-networks/fifa}.

\bibitem{SEIERSTAD201144}
C.~Seierstad and T.~Opsahl, For the few not the many? {T}he effects of
  affirmative action on presence, prominence, and social capital of women
  directors in {N}orway. \textit{Scandinavian Journal of Management}
  \textbf{27}, 44--54 (2011).

\bibitem{MFB15}
R.~Mastrandrea, J.~Fournet, and A.~Barrat, Contact patterns in a high school: A
  comparison between data collected using wearable sensors, contact diaries and
  friendship surveys. \textit{PLOS One} \textbf{10}, e107878 (2015).

\bibitem{CF08}
N.~A. Cristakis and J.~H. Fowler, The collective dynamics of smoking in a large
  social network. \textit{New England Journal of Medicine} \textbf{358},
  2249--2258 (2008).

\bibitem{kross2013facebook}
E.~Kross, P.~Verduyn, E.~Demiralp, J.~Park, D.~S. Lee, N.~Lin, H.~Shablack,
  J.~Jonides, and O.~Ybarra, Facebook use predicts declines in subjective
  well-being in young adults. \textit{PLOS One} \textbf{8}, e69841 (2013).

\bibitem{LYW16}
K.~Lerman, X.~Yan, and X.-Z. Wu, The majority illusion in social networks.
  \textit{PLOS One} \textbf{11}, e147616 (2016).

\bibitem{nettasinghe2019your}
B.~Nettasinghe and V.~Krishnamurthy, What do your friends think?  Efficient
  polling methods for networks using friendship paradox. \textit{IEEE
  Transactions on Knowledge and Data Engineering}  (2019).

\bibitem{rothschild2009forecasting}
D.~Rothschild, Forecasting elections: Comparing prediction markets, polls, and
  their biases. \textit{Public Opinion Quarterly} \textbf{73}, 895--916
  (2009).

\end{thebibliography}
\end{document}